\documentclass[preprint,aps,amsmath,nofootinbib,floatfix,superscriptaddress]{revtex4-1}

\usepackage{graphicx}
\usepackage{epstopdf}
\usepackage{subfigure}
\usepackage{multirow}

\usepackage{slashed}

\usepackage{color}
\usepackage{comment}
\usepackage{booktabs} 
\usepackage{natbib}
\usepackage[colorlinks,linkcolor=black,anchorcolor=blue,citecolor=green]{hyperref}
\makeatletter

\newcommand{\Rmnum}[1]{\expandafter\@slowromancap\romannumeral #1@}
\makeatother

\begin{document}
\title{Testing the Nature of Neutrino from Four-Body $\tau$ Decays}
\author{Han Yuan}
\email{hanyuan@hit.edu.cn}
\affiliation{Department of Physics, Harbin Institute of Technology, Harbin, 150001, China}

\author{Yue Jiang}
\email{jiangure@hit.edu.cn}
\altaffiliation[]{Corresponding author}
\affiliation{Department of Physics, Harbin Institute of Technology, Harbin, 150001, China}

\author{Tian-hong Wang}
\email{thwang@hit.edu.cn}
\affiliation{Department of Physics, Harbin Institute of Technology, Harbin, 150001, China}

\author{Qiang Li}
\email{lrhit@protonmail.com}
\affiliation{Department of Physics, Harbin Institute of Technology, Harbin, 150001, China}

\author{Guo-Li Wang}
\email{gl\_wang@hit.edu.cn}
\altaffiliation[]{Corresponding author}
\affiliation{Department of Physics, Harbin Institute of Technology, Harbin, 150001, China}
\begin{abstract}
This paper is designed to discuss four-body lepton number violating tau decay. We study the processes $\tau^+ \to e^+ e^+ \pi^- \bar{\nu}_\tau$ and $\tau^+ \to e^+ e^+ \pi^- \nu_e$ to determine the nature of neutrino. The first process violates lepton number by two units which can only happen through a internal Majorana. The second one conserves lepton number but violates lepton flavor which can take place with both Majorana neutrino and Dirac neutrino. We calculate their branching ratio $Br$ and differential branching ratio $dBr/dE_\pi$ to distinguish the Majorana neutrino vs. Dirac neutrino. We also propose the possibility of experiment to perform this detection.
\end{abstract}
\maketitle

\section{Introduction}

The existence of neutrino mass has been demonstrated by many neutrino experiments \cite{Fukuda1998,Eguchi2003,Ahmed2004,Argyriades2009,Wendell2010}. Furthermore, the mixing angles in neutrino oscillation have been detected\cite{dayabay}, which also means that neutrinos have masses. In this perspective, Standard Model (SM) should be expanded since {in SM the neutrino is massless and only has left hand state (or only right hand state for anti-neutrino)}. Certainly there are many ways to expand SM and theoretically explain the neutrino mass such as supersymmetric (SUSY)\cite{Giang2012,Hue2013}, see-saw model\cite{Dinh2013} and extended $SU(3)_c\times SU(3)_L\times U(1)_x$ (331) models\cite{Fonseca2016}. However before expanding SM, we still have fundamental questions about neutrino physics. All the other fermions of SM are Dirac ones, but we are still not sure whether the neutrino is a Majorana \cite{Majorana:1937vz} neutrino or a Dirac neutrino.

Majorana nature of neutrino is attractive, since Majorana neutrino and its anti-particle are the same that can cause $|\Delta L|=2$ Lepton Number Violating (LNV) decays. This process is forbidden for Dirac neutrino, so it can be regarded as one method to experimentally demonstrate the nature of neutrino. The existence of heavy, mostly-sterile neutrino is also interesting, which can be a candidate for the dark matter\cite{Dodelson:1993je}, explain the supernova explosion\cite{Fuller:2009zz}, account for the baryogenesis\cite{Fukugita:1986hr} and leptogenesis\cite{Buchmuller:2005eh}, etc.

There are many kinds of $|\Delta L|=2$ processes. The neutrinoless double beta decays ($0\nu\beta\beta$) in nuclei are regarded as the most sensitive way\cite{Racah:1937qq,Furry:1939qr,Doi:1985dx}. { The neutrinoless double beta decays ($0\nu\beta\beta$) in nuclei are regarded as the most sensitive way. Finding these decay showing that the neutrino is Majorana neutrino and the LNV process of nuclei can also provide the information about heavy neutrino mixing with charged leptons. But writing the nuclear matrix element is still a difficult task in theory which may cause difficulty in calculation of the $0\nu\beta\beta$ decay of nuclei. Even though, the neutrinoless double beta decay of nuclei can also put stringent bounds on the heavy neutrinos.} Some other ways are the heavy meson decay $M_1\rightarrow M_2\ell\ell$  \cite{Seon2011,Lees2012,Aaij2014}, various tau decays  \cite{Miyazaki2013,Miyazaki2010,LHCbcollaboration2013,Hayasaka2010} and $pp$ collisions with final $\mu^\pm\mu^\pm$ and $e^\pm e^\pm$ jets\cite{Chatrchyan2012}. Along with the energy enhanced in the LHC, LNV decays of the Higgs boson have the possibility to be discovered\cite{Maiezza2015}. Furthermore Ref. \cite{Peng2016} analyzes the sensitivity of next-generation tonne-scale neutrinoless double $\beta$-decay experiments and searches for like sign di-electrons plus jets at the LHC to TeV scale lepton number violating interactions. Sometimes baryon number violating is also connected with lepton number violating \cite{LHCbcollaboration2013,McCracken2015}. Meson rare decays where $|\Delta L|=2$ such as three body meson decays $M_1^+\rightarrow\ell_1^+\ell_2^+M_2^-$ have been studied in Refs.  \cite{Littenberg1992,Littenberg2000,Atre2009,Cvetic2010,Zhang2011,Chen2012,Milanes:2016rzr} and four-body decays ${M}\rightarrow M_1\ell\ell M_2$ in Refs. \cite{Delepine2011,Yuan2013,Castro:2013jsn,Milanes:2016rzr} have also been calculated seriously.

Besides the upper processes, there are also some other results about LNV (LFV) processes in experiment. Belle reports its result about detecting such decays with 719 million produced $\tau^+\tau^-$ pairs in Ref.  \cite{Hayasaka2010}. LHCb searches for such decays at $\sqrt s=7$ TeV in Ref. \cite{LHCbcollaboration2013}. Both Belle and LHCb show that $\textit{B}(\tau^-\rightarrow\mu^-\mu^+\mu^-)<10^{-8}$. Ref. \cite{Miyazaki2010} reportes the upper limits on the  branching ratios of $|\Delta L|=2$ tau decay $\tau^-\rightarrow\ell^+M_1^-M_2^-$ in the order of $10^{-8}$. In Ref. \cite{Castro2012} most tau four-body decays like $\tau^-\rightarrow\nu_\tau\ell_1^-\ell_2^-M^+$ have the branching ratios close to the order of $10^{-7}$. However among the decays calculated in theory, Ref. \cite{Castro2012} suggests the largest branching fraction shown in $\tau^-\rightarrow\bar{\nu}_\tau e^-e^-\pi^+$ approaches $10^{-4}$. This result is very impactive and motivated for us to do more theoretical calculation in $|\Delta L|=2$ four-body tau decay.

Specially we consider the LNV decay $\tau^+ \to e^+ e^+ \pi^- \bar{\nu}_\tau$ which is induced by exchanging Majorana neutrino. However, the final active light neutrino $\bar{\nu}_\tau$ is missing energy in experiment whose flavor is not sure. So a similar lepton number conserve, but LFV process $\tau^+ \to e^+ e^+ \pi^- \nu_e$ should be added into consideration, which is induced by exchanging either Majorana neutrino or Dirac neutrino. For $\tau^+ \to e^+ e^+ \pi^- \bar{\nu}_\tau$ process, the final neutrino is $\bar{\nu}_\tau$ produced at the $\tau^+$ vertex. As the reason of PMNs mixing, the final anti-neutrino's flavor is not fixed, which can be any one. However, at the very moment when the final anti-neutrino has just produced, it must be $\bar{\nu}_\tau$ as the initial lepton is $\tau$. With propagating distance of the resonance lengthen, the anti-neutrino $\bar{\nu}_\tau$ can change to any other flavor. But in this work, we do not consider the final state of the neutrino/anti-neutrino resonance flavor, since it is missing energy in experiments, after all. Similarly, in $\tau^+ \to e^+ e^+ \pi^- \nu_e$, the final neutrino is $\nu_e$ produced at the $e^+$ vertex. Due to the reason stated above, we also do not consider the PMNs mixing of this process, too. And we will introduce how to distinguish these two processes to determine the nature of neutrino.

In our calculation, we choose the previously used phenomenology model \cite{Atre2009}, where there are $3+n$ generation Majorana neutrinos. The first $3$ generations are light active neutrinos, and the other $n$ generations are heavy sterile neutrinos. The concrete parameters including mixing parameters and neutrino masses should be determined by experiments. As well known, the rare decays induced by virtual neutrino are suppressed heavily either by the factor $\frac{m_{\nu\ell}^2}{m^2_{W}}$ (if it is a light active neutrino) or by the mixing parameter $V^2_{\ell N}$ (if it is a heavy sterile neutrino), so these processes are hard to be detected by current experiments. But if the exchanging neutrino is on mass-shell, the corresponding decay rate will be enhanced several orders larger \cite{Atre2009,Zhang2011,Yuan2013}, which may be reached by current experiments, so we will only focus on the neutrino-resonance processes. Since the available phase space range of the exchanging neutrino mass is $m_{\pi}\lesssim m_N\lesssim m_{\tau}$ in the decay $\tau^+ \to e^+ e^+ \pi^- \nu$ (the final state $\nu$, $\nu_{e}$ or $\nu_{\tau}$ is light active neutrino), so it is heavy, and it should be a fourth generation sterile neutrino.

In section 2 we show the calculation of the $\tau^+ \to e^+ e^+ \pi^- \bar{\nu}_\tau$ and $\tau^+ \to e^+ e^+ \pi^- \nu_e$. Then we display the result and analysis it in section 3 in particular discuss how to distinguish the Majorana neutrino and Dirac neutrino through the processes mentioned above. Lastly, in section 4 we give our conclusion.

\section{Calculation details of $\tau^+ \to e^+ e^+ \pi^- \bar{\nu}_\tau$ and $\tau^+ \to e^+ e^+ \pi^- \nu_e$}

\begin{figure}[!h]
\centering
\includegraphics[scale=0.4]{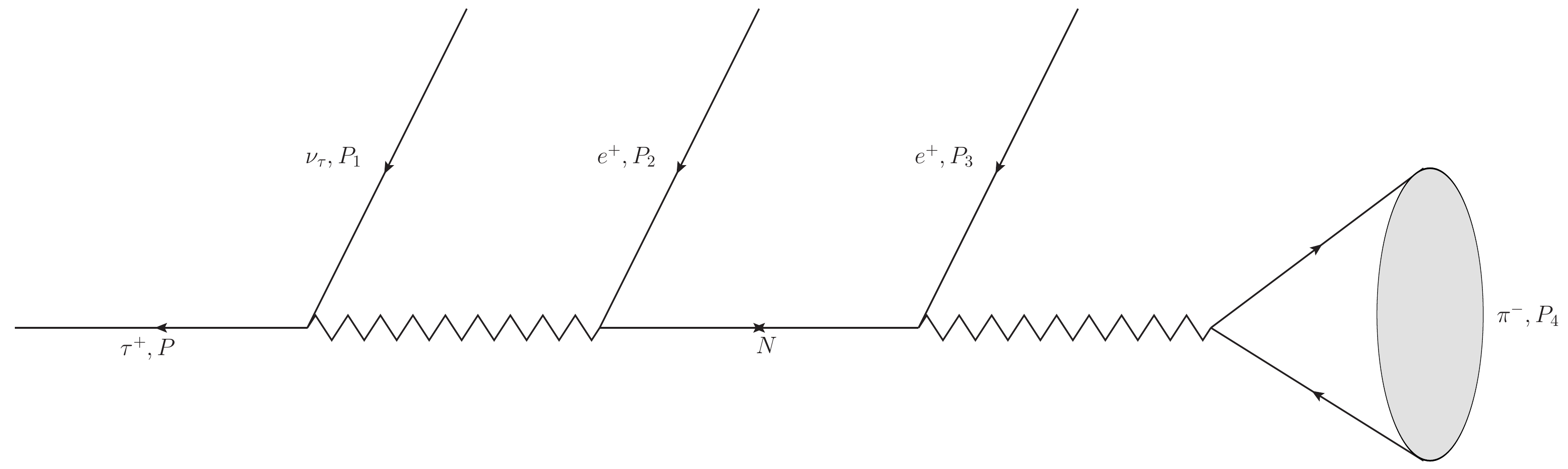}
\caption{Feynman diagram of the LNV process $\tau^+ \to e^+ e^+ \pi^- \bar{\nu}_\tau$.\label{fig:feynM}}
\end{figure}

The feynman diagram of $\tau^+ \to e^+ e^+ \pi^- \bar{\nu}_\tau$ is shown in figure~\ref{fig:feynM},
where $\tau$ with momentum $P$, neutrino $\nu_{\tau}$ with momentum $P_1$, two electron with momentum $P_2$ and $P_3$ and meson $\pi$ with momentum $P_4$. As we know the decay $\tau^+ \to e^+ e^+ \pi^- \bar{\nu}_\tau$ is forbidden under the frame of SM. Because it violates the Lepton number by two units. However if the propagater in this process is not traditional type neutrino but Majorana neutrino, this processe may occur. Following previous studies \cite{Atre2009,Bar-Shalom2006}, the charged current interaction lagrangian for this $|\Delta L|=2$ decay in terms of neutrino mass eigenstates is
\begin{equation}
\mathcal{L}_{cc}=-\frac{g}{\sqrt{2}}W^+_{\mu}\left(\sum_{\ell=e}^{\tau}\sum^3_{m=1}U^{l\nu*}_{\ell m}\bar{\nu}_m\gamma^{\mu}P_L\ell+\sum^{\tau}_{\ell=e}\sum^{3+n}_{m^\prime=4}V_{\ell m^\prime}^{lN*}\bar{N}^c_{m^\prime}\gamma^{\mu}P_L\ell\right)+\mathrm{h.c.},
\end{equation}
where $\ell=e, \mu, \tau$, $U_{\ell m}$ and $V_{\ell m^\prime}$ are mixing matrices, $\nu_m$ and $N_{m^\prime}$ are neutrino mass eigenstates and $P_L=\frac{1}{2}(1-\gamma_5)$. There are $3+n$ generation neutrinos: when $m=1, 2, 3$ they are active neutrinos whose masses $m_{\nu_m}\sim \mathcal{O}(\mathrm{eV})$\cite{Seljak:2004xh} and mixing parameter $U^{l\nu*}_{\ell m}$ is large \cite{Atre2009}. When $m\geq4$, they are heavy sterile neutrinos whose masses $m_{N_m^\prime}\sim\mathcal{O}(\mathrm{MeV}-\mathrm{GeV})$ and mixing parameter $V_{\ell m^\prime}^{lN*}$ is small. Considering Fig.~\ref{fig:feynM}, the light neutrino propagator's contribution will be suppressed by the small neutrino mass $\frac{m^2_{\nu_m}}{M_W^2}$. So we drop the $\sum_{m=1}^3$ part. In principle all the heavy Majorana neutrinos will contribute to the amplitude. But for simplicity, only one heavy neutrino is considered. So we only consider the lightest heavy Majorana neutrino which should be the fourth generation if exist. Thus the considered charged current interaction lagrangian can be rewritten as
\begin{equation}
\mathcal{L}_{cc}=-\frac{g}{\sqrt{2}}W^+_{\mu}\sum^{\tau}_{\ell=e}V_{\ell4}^*\bar{N}^c_4\gamma^{\mu}P_L\ell+\mathrm{h.c.}.
\end{equation}

From the lagrangian we can get the propagator of heavy Majorana neutrino
\begin{equation}
\frac{\slashed{q}+m_4}{q^2-m_4^2+i\Gamma_{N_4}m_4},
\end{equation}
where $q$ is the momentum of heavy Majorana neutrino and $\Gamma_{N_4}$ is the decay width of heavy Majorana neutrino. We can see the heavy Majorana neutrino contribution has a resonant enhancement when $q^2\approx m_4^2$, so from now on we choose it on mass shell.

The amplitude of $\tau^+ \to e^+ e^+ \pi^- \bar{\nu}_\tau$ can be written as
\begin{equation}\label{am1}
\mathcal{M}=\frac{g^2 V_{ud}}{8M_W^4}\bar{v}(P)\gamma^\mu (1-\gamma_5)v(P_1)\times \mathcal{L}_{\mu\nu}\times \langle \pi(P_4)|{\bar{q}_1}\gamma^\nu(1-\gamma_5)q_2|0\rangle,
\end{equation}
where the momentum dependence in the propagator of $W$ boson has been ignored since it is much smaller than the $W$ mass; $g$ is the weak coupling constant; $V_{ud}$ is the Cabibbo-Kobayashi-Maskawa (CKM) matrix element between quarks $\bar u$ and $d$ in $\pi$; $\mathcal{L}_{\mu\nu}$ is the transition amplitude of the leptonic part.

\begin{figure}[!h]
\centering
\includegraphics[scale=0.5]{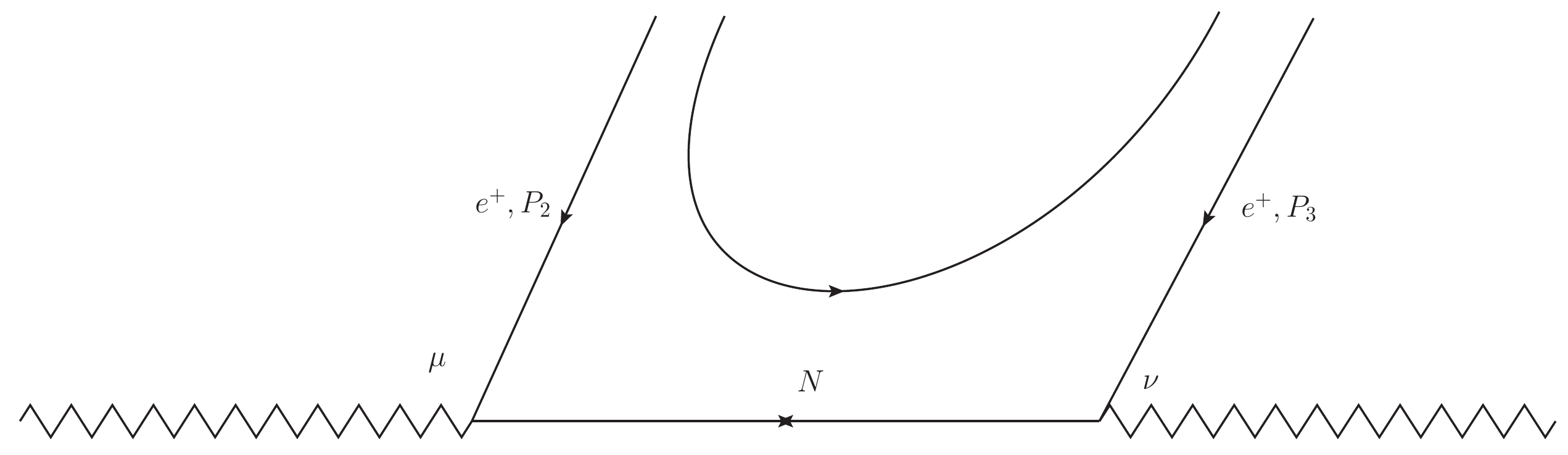}
\caption{Feynman diagram for the $|\Delta L|=2$ leptonic part of the LNV process $\tau^+ \to e^+ e^+ \pi^- \bar{\nu}_\tau$.\label{fig:LV2}}
\end{figure}

The $|\Delta L|=2$ leptonic part can be separated from the whole process, and the feynman diagram for this part is shown in Fig. \ref{fig:LV2}.
We follow Ref. \cite{Denner:1992vza} to write the amplitude of this part. First we need to draw a fermion flow line. This fermion flow starts from $e_1^+$ and points to $e_2^+$.  The specific situation can be found in Fig. \ref{fig:LV2}. Then we write from an external leg proceeding opposite to the chosen orientation (fermion flow) through the chain. The amplitude is
\begin{equation}\label{eq:DeltaL}
\mathcal{L_{\mu\nu}}=\frac{\mathit g^2}{2}|V_{e4}|^2m_4\frac{{\bar{u}(P_3)}\gamma_{\mu}\gamma_{\nu} P_Lv(P_2)}{q^2-m_4^2+i\Gamma_{N_4}m_4}.
\end{equation}

The third term in the right side of Eq. (4) can be described with the decay constant of meson $\pi$. As $\pi$ is a pseudoscalar, the amplitude can be written as
\begin{equation}\label{eq:constant}
\langle \pi(P_4)|{\bar{q}_1}\gamma^\nu(1-\gamma_5)q_2|0\rangle=i F_{\pi}P_4^\nu,
\end{equation}
where $F_\pi$ is the decay constant of $\pi$.

Combining Eq. (\ref{eq:DeltaL}) and Eq. (\ref{eq:constant}),  Eq. (4) can be written as
\begin{eqnarray}\label{eq:amplitudeM}
\mathcal{M}=2G_F^2|V_{e4}|^2V_{ud} F_\pi m_4{\bar{v}(P)}\gamma_\mu(1-\gamma_5)v(P_1)\bar{u}(P_3){{\frac{\gamma^\mu\slashed{P_4}}{(P_3+P_4)^2-m^2_4+i{\Gamma}_{N}m_4}}}P_Lv(P_2),
\end{eqnarray}
where $G_F$ is Fermi constant.

\begin{figure}[!h]
\centering
\includegraphics[scale=0.4]{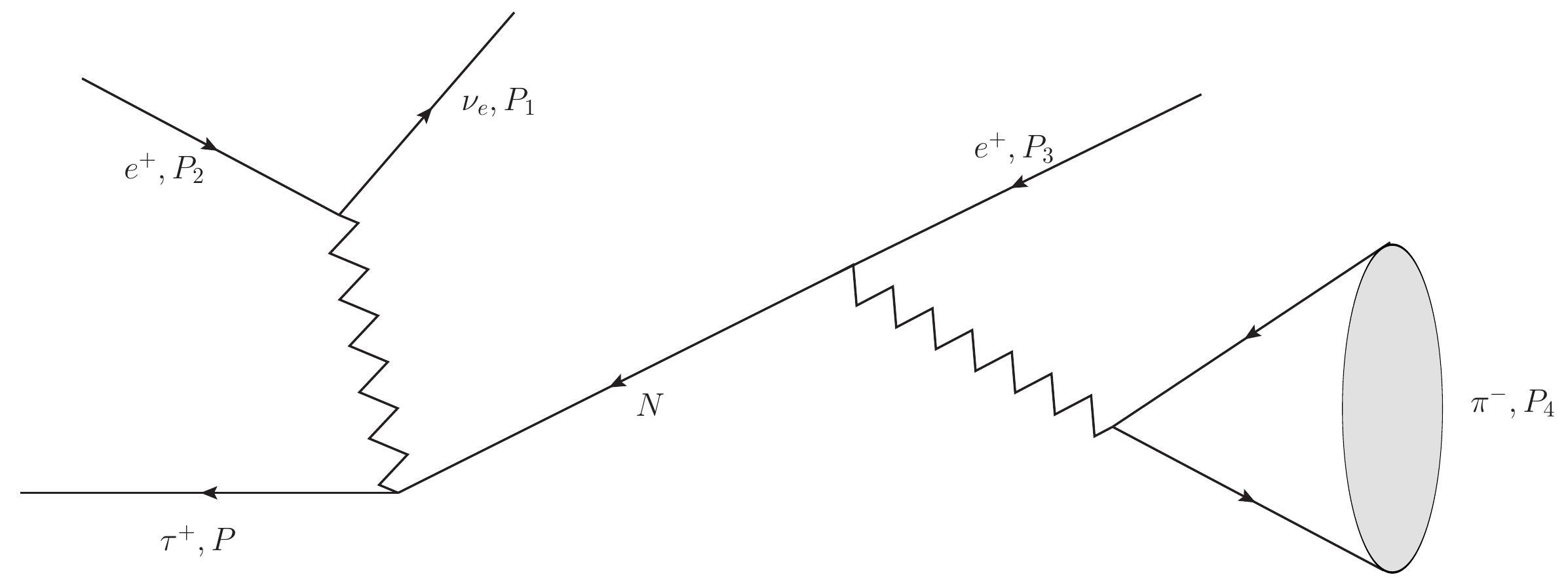}
\caption{Feynman diagram of the LFV process $\tau^+ \to e^+ e^+ \pi^- \nu_e$.\label{fig:feynMD}}
\end{figure}

As discussed above, process $\tau^+ \to e^+ e^+ \pi^- \nu_e$ should be added into consideration. The Feynman diagram is drawn in Fig. \ref{fig:feynMD},
where the neutrino propagator can be a Majorana or a Dirac neutrino. Analysising these two processes we found that if $N$ is Majorana neutrino these two processes (Fig. 1 and Fig. 3) should occur, otherwise only the second one (Fig. 3) is possible for Dirac neutrino. So for $\tau^+ \to e^+ e^+ \pi^- \nu_e$ with Majorana neutrino we use the normal feynman rules in SM to write the amplitude
\begin{eqnarray}\label{eq:amplitudeD}
\mathcal{M}=2G_F^2V_{e4}V_{\tau4}V_{ud} F_\pi {\bar{u}(P_1)}\gamma_\mu(1-\gamma_5)v(P_2)\bar{v}(P)\gamma^\mu{{\frac{\slashed{q}\slashed{P_4}}{q^2-m^2_4+i{\Gamma}_{N}m_4}}}P_Lv(P_3).
\end{eqnarray}
Thus whether the propagator is Dirac neutrino or Majorana neutrino, the amplitudes are the same.

Considering resonant enhancement, we choose the intermediate neutrino on mass-shell with enough life time. Then the process $\tau^+ \to e^+ e^+ \pi^- \bar{\nu}_\tau$ can be separated into two parts, $\tau^+ \to e^+ \bar{\nu}_\tau N$ and $N\rightarrow e^+ \pi^- $. So the two electrons are different and can be distinguished by vertexes in experiment. So we do not need to consider the exchange of them. And for heavy neutrino its mass is much larger than its decay width (it is a sterile neutrino with only weak interaction), so we can use Narrow Width Approximation (NWA) to simplify the phase space integration \cite{Atre2009,Zhang2011}. Relevant calculation details can be found in the appendix.

\section{Results and analysis}
There are some important input parameters in our calculation, such as the decay width of neutrino and the mixing parameters between charged lepton and neutrino. In order to get the decay width of heavy neutrino we follow the method in Ref. \cite{Atre2009}, which calculated all the possible decay modes of Majorana neutrino to get its witdh $\Gamma_{N_4}$. The Majorana neutrino decay channels will include charge-conjugate channels, because its antiparticle is the same as itself. Therefore the width $\Gamma_{N_4}$ is twice as large as Dirac neutrino. Thus we can get the decay width $\Gamma_{N_4}$ of Dirac neutrino. Regarding our choice of neutrino mixing parameters, we follow the Ref. \cite{delAguila:2008pw} to choose the parameters $|V_{eN}|^2=3.0\times10^{-3}$ and $|V_{\tau N}|^2=6.0\times10^{-3}$. In Ref. \cite{Atre2009} the limit of $|V_{eN}|^{2}$ in the mass range 0.5 - 1.6 GeV is $|V_{eN}|^{2}\sim10^{-5}$. In Ref. \cite{Faessler:2014kka} the more stronger limit is $|V_{eN}|^{2}\sim10^{-8}$. And the limits about Dirac neutrino mixing parameters can indeed be abstracted from lepton flavour violating processes like $\tau^{-} \rightarrow l^{-}l^{+}l^{-}$. Refs. \cite{LHCbcollaboration2013,Miyazaki2010} show the branching ratios of $\tau\to3\ell$ processes are less than $10^{-8}$. It can reflect the the mixing parameters, to which the branching ratios is in direct proportion, is smaller than $10^{-4}$ \cite{Hue:2013uw}. This paper aims to show the differences between $\tau^+ \to e^+ e^+ \pi^- \bar{\nu}_\tau$ and $\tau^+ \to e^+ e^+ \pi^- \nu_e$, which represent the distinction between Majorana and Dirac neutrino. As the mixing parameters influence branching ratio obviously, we should reduce the influence from these parameters and focus on the two processes themselves. Under these circumstances, we first choose $|V_{eN}|^2=3.0\times10^{-3}$ and $|V_{\tau N}|^2=6.0\times10^{-3}$ to obtain the branching ratios and differential branching ratios of $\tau^+ \to e^+ e^+ \pi^- \bar{\nu}_\tau$ and $\tau^+ \to e^+ e^+ \pi^- \nu_e$, since there are no exact value of the mixing parameters between heavy neutrino and charged lepton. Then we try to get the differential branching ratio with little influence of mixing parameters.

\begin{figure}[!h]
\centering
\includegraphics[scale=0.5]{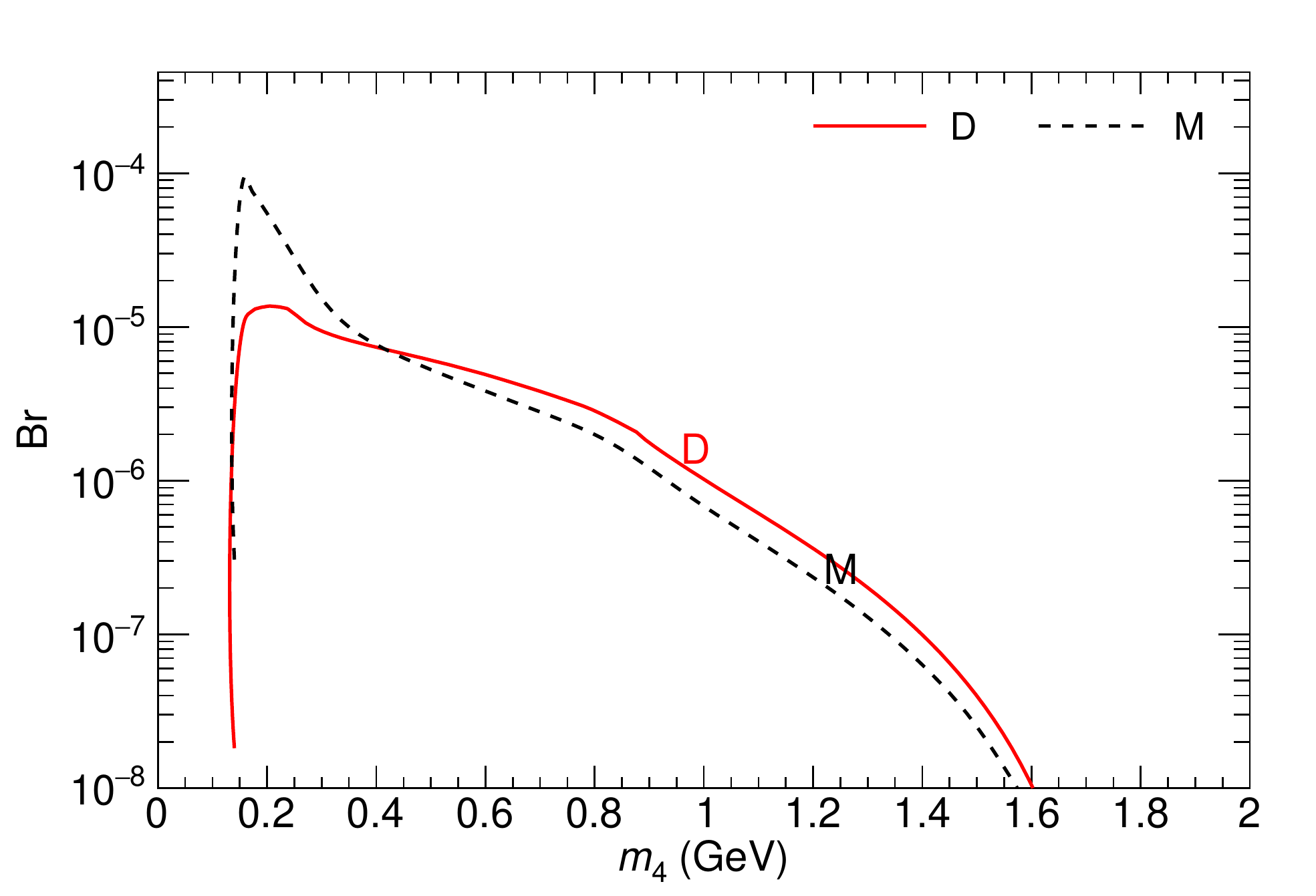}
\caption{Branching ratio as a function of the heavy neutrino mass.\label{fig:tot}}
\end{figure}

We are only interested in the processes when the exchanging neutrinos are on mass shell, but we do not calculate all the possible cases available by the phase space. In this research we choose several masses in the possible kinematics mass range such as $0.5$, $0.8$, $1.0$, $1.2$, $1.4$ and $1.6~\mathrm{GeV}$ to get the results. We derive the branching ratios of $\tau^+ \to e^+ e^+ \pi^- \nu$ ($\bar{\nu}_\tau$ or $\nu_e$) and $\tau^+ \to e^+ e^+ \pi^- \nu_e$. They are shown in Tab. \ref{tab:Br}.
\begin{table}[ph]
\caption{Branching ratio of $|\Delta L|=2$ process $\tau^+ \to e^+ e^+ \pi^- \nu$ and $\tau^+ \to e^+ e^+ \pi^- \nu_e$ with $|V_{eN}|^2=3.0\times10^{-3}$ and $|V_{\tau N}|^2=6.0\times10^{-3}$}
\begin{center}
{\begin{tabular}{|c|c|c|} \hline
$m_4~[\mathrm{GeV}]$&$Br^{(M)}~(\tau^+ \to e^+ e^+ \pi^- \bar{\nu}_\tau)$ and $Br^{(M)}~(\tau^+ \to e^+ e^+ \pi^- \nu_e)$ &$Br^{(D)}~(\tau^+ \to e^+ e^+ \pi^- \nu_e)$\\\hline
$0.5$&$5.27\times10^{-6}$&$6.07\times10^{-6}$\\\hline
$0.8$&$2.05\times10^{-6}$&$2.90\times10^{-6}$\\\hline
$1.0$&$7.17\times10^{-7}$&$9.83\times10^{-7}$\\\hline
$1.2$&$2.30\times10^{-7}$&$3.55\times10^{-7}$\\\hline
$1.4$&$6.33\times10^{-8}$&$9.94\times10^{-8}$\\\hline
$1.6$&$7.03\times10^{-9}$&$1.11\times10^{-8}$\\\hline
\end{tabular} \label{tab:Br}}
\end{center}
\end{table}
Fig.\ref{fig:tot} shows branching ratio as a function of the heavy neutrino mass.
The black dash line represents the branching ratio of process $\tau^+ \to e^+ e^+ \pi^- \nu$ ($\bar{\nu}_\tau$ or $\nu_e$) with internal Majorana neutrino and the red solid line is about $\tau^+ \to e^+ e^+ \pi^- \nu_e$ with internal Dirac neutrino. In our calculation the total branching ratio with Majorana neurino is the sum of $\tau^+ \to e^+ e^+ \pi^- \bar{\nu}_\tau$ and $\tau^+ \to e^+ e^+ \pi^- \nu_e$. If the internal neutrino is Dirac neutrino, the total branching ratio is double times of $\tau^+ \to e^+ e^+ \pi^- \nu_e$, since the decay width $\Gamma_N$ in Eq.\eqref{eq:amplitudeD} of Dirac neutrino is half of Majorana neutrino.
From Tab.\ref{tab:Br} we can see that, with the mixing parameters $|V_{eN}|^2=3.0\times10^{-3}$ and $|V_{\tau N}|^2=6.0\times10^{-3}$, the branching ratio of Dirac neutrino is much larger than that of the internal Majorana neutrino. The branching ratio of $\tau^+ \to e^+ e^+ \pi^- \bar{\nu}_\tau$ and $\tau^+ \to e^+ e^+ \pi^- \nu_e$ should be roughly equal. The reason is both two processes can be separated into two sub-processes: one is three-body $\tau^+\to\ell^+\nu N$ and the other is a secondary two-body process $N\to\ell^+\pi^-$. The branching ratio of $\tau^+ \to e^+ e^+ \pi^- \bar{\nu}_\tau$ is noted as $Br_1$ while the branching ratio of $\tau^+ \to e^+ e^+ \pi^- \nu_e$ is noted as $Br_2$. The principal cause which can results in the differences between $Br_1$ and $Br_2$ is mixing parameter. For $\tau^+ \to e^+ e^+ \pi^- \bar{\nu}_\tau$ the mixing parameter is $|V_{eN}|^2=3.0\times10^{-3}$ and for $\tau^+ \to e^+ e^+ \pi^- \nu_e$ it is $|V_{\tau N}|^2=6.0\times10^{-3}$. Thus $Br_2$ is about twice of $Br_1$. For Majorana case the total branching ratio is the sum of $Br_1$ and $Br_2$, while in Dirac case it is $2\times Br_2$. Tab.\ref{tab:Brr} shows the branching ratio with $|V_{eN}|^2=3.0\times10^{-3}$ and $|V_{\tau N}|^2=3.0\times10^{-3}$.
\begin{table}[ph]
\caption{Branching ratio of $|\Delta L|=2$ process $\tau^+ \to e^+ e^+ \pi^- \nu$ and $\tau^+ \to e^+ e^+ \pi^- \nu_e$ with $|V_{eN}|^2=3.0\times10^{-3}$ and $|V_{\tau N}|^2=3.0\times10^{-3}$}
\begin{center}
{\begin{tabular}{|c|c|c|} \hline
$m_4~[\mathrm{GeV}]$&$Br^{(M)}~(\tau^+ \to e^+ e^+ \pi^- \bar{\nu}_\tau)$ and $Br^{(M)}~(\tau^+ \to e^+ e^+ \pi^- \nu_e)$ &$Br^{(D)}~(\tau^+ \to e^+ e^+ \pi^- \nu_e)$\\\hline
$0.5$&$3.75\times10^{-6}$&$3.03\times10^{-6}$\\\hline
$0.8$&$1.32\times10^{-6}$&$1.45\times10^{-6}$\\\hline
$1.0$&$4.10\times10^{-7}$&$4.92\times10^{-7}$\\\hline
$1.2$&$1.41\times10^{-7}$&$1.78\times10^{-7}$\\\hline
$1.4$&$3.85\times10^{-8}$&$4.97\times10^{-8}$\\\hline
$1.6$&$4.24\times10^{-9}$&$5.58\times10^{-9}$\\\hline
\end{tabular} \label{tab:Brr}}
\end{center}
\end{table}
Reducing the influence from mixing parameters, we can see the branching ratios of Majorana case and Dirac case are similar at this time. Another cause may bring difference to the branching ratios of these two processes is the leptonic tensor part. In Eq. \eqref{eq:amplitudeM} Feynman rules of the two vertexes corresponding to the two charged leptons are different, but in Eq. \eqref{eq:amplitudeD} they are the same. So, under the effect of vertex factors, the numerator $(m_4+\slashed{q})$ of the propagator left is merely $m_4$ in Eq. \eqref{eq:amplitudeM} and $\slashed{q}$ in Eq. \eqref{eq:amplitudeD}. Eq. \eqref{eq:lv2width} shows the decay width of $\tau^+ \to e^+ e^+ \pi^-\bar{\nu}_\tau\propto\frac{1}{m_4}$, thus as the heavy neutrino mass growing, the branching ratio gets smaller. So is the process $\tau^+ \to e^+ e^+ \pi^-{\nu}_e$. But since we do not know the exact value of mixing parameter, the total branching ratio cannot be used to distinguish the $\tau^+ \to e^+ e^+ \pi^- \nu$ and $\tau^+ \to e^+ e^+ \pi^- \nu_e$.

\begin{figure}[ht] \label{result}
\centering
\subfigure[]{
   \includegraphics[scale=0.35] {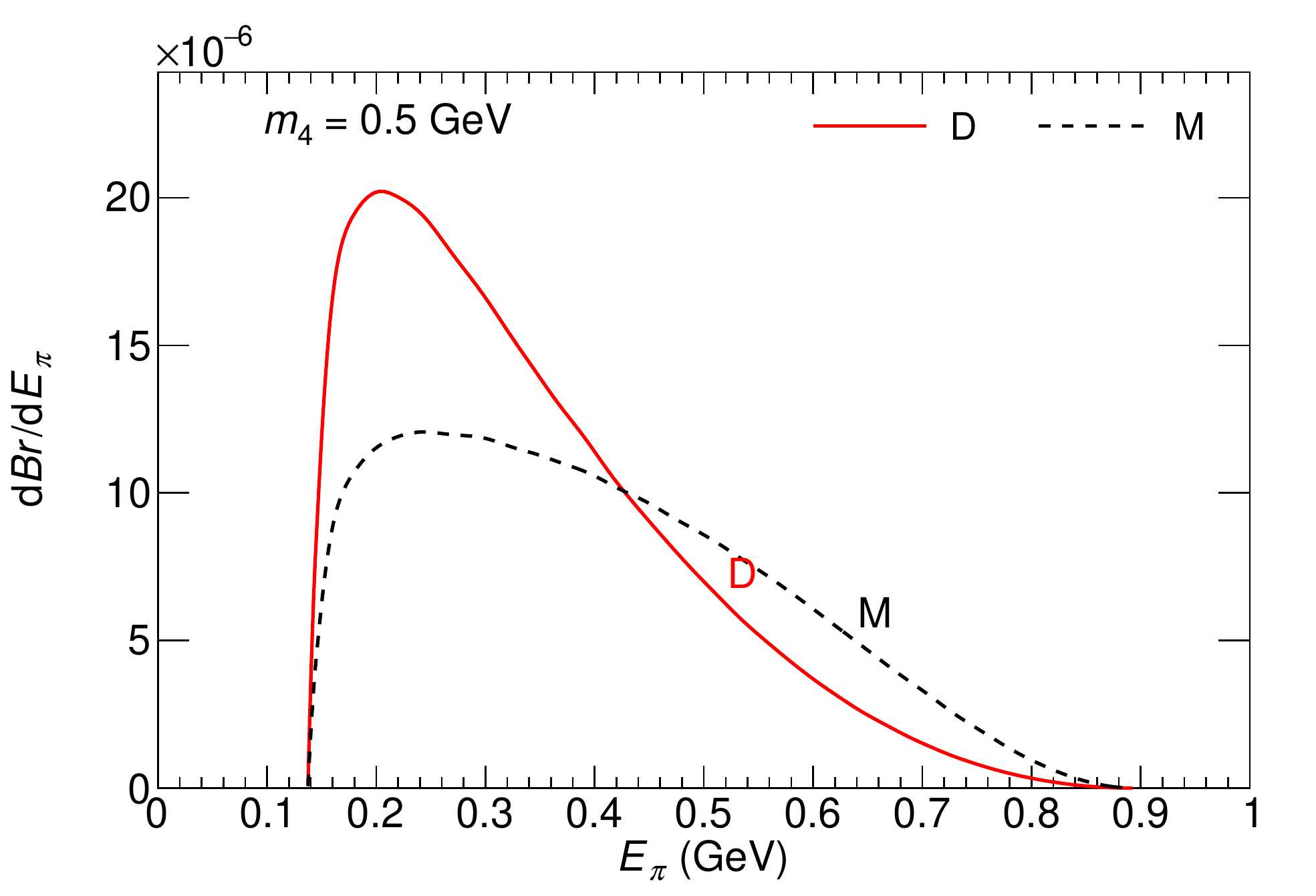}
   \label{a}
 }
 \subfigure[]{
   \includegraphics[scale=0.35] {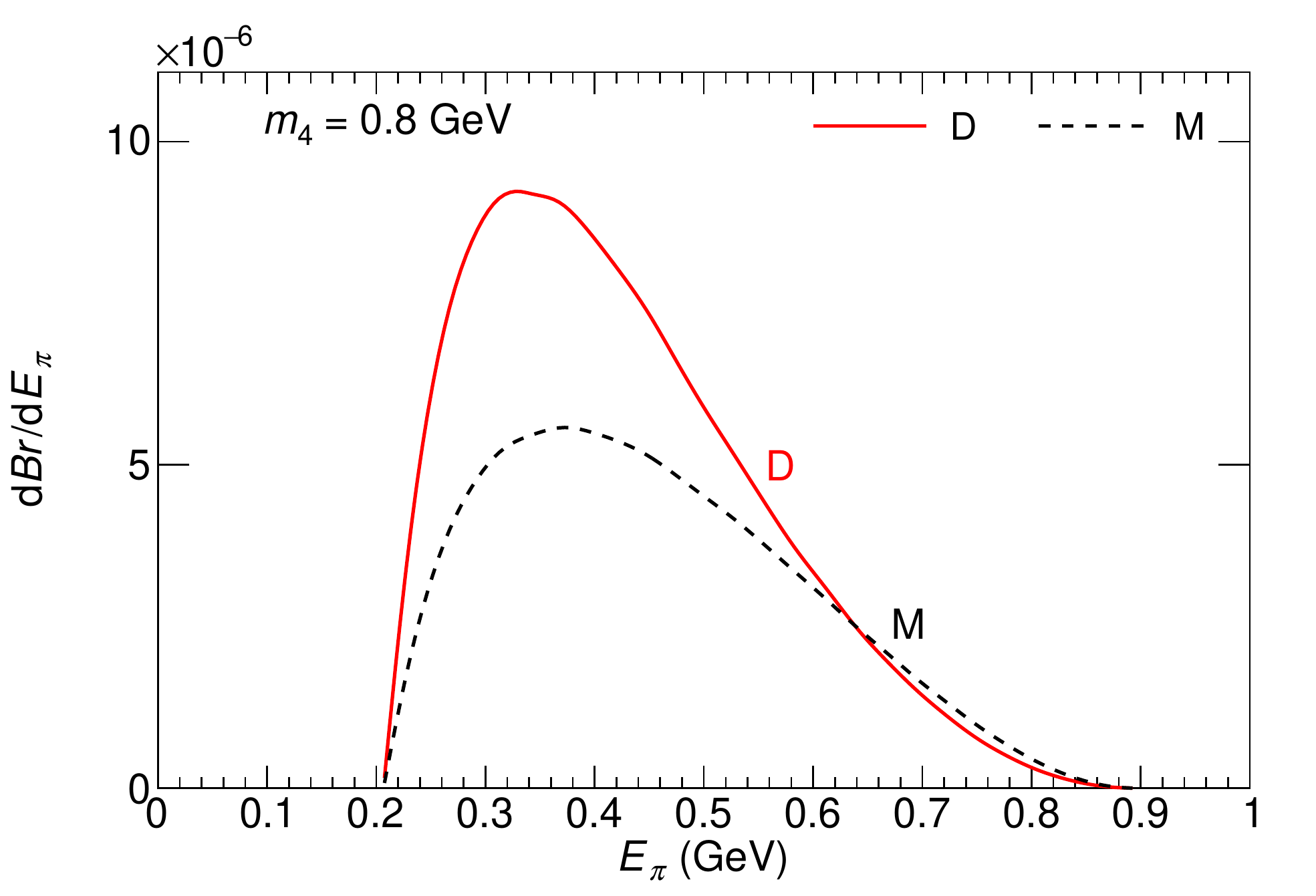}
   \label{b}
 }
 \subfigure[]{
   \includegraphics[scale=0.35] {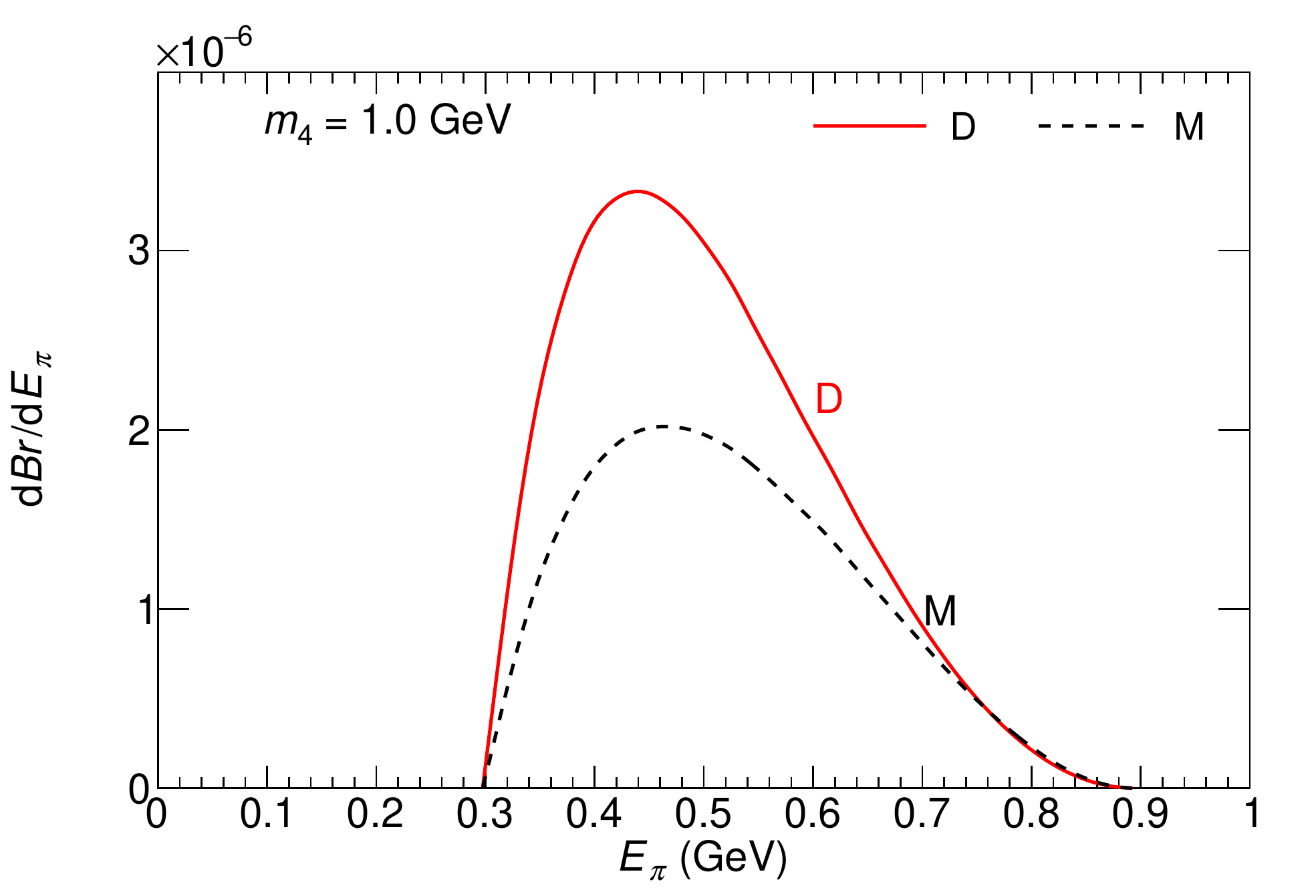}
   \label{c}
 }
  \subfigure[]{
   \includegraphics[scale=0.35] {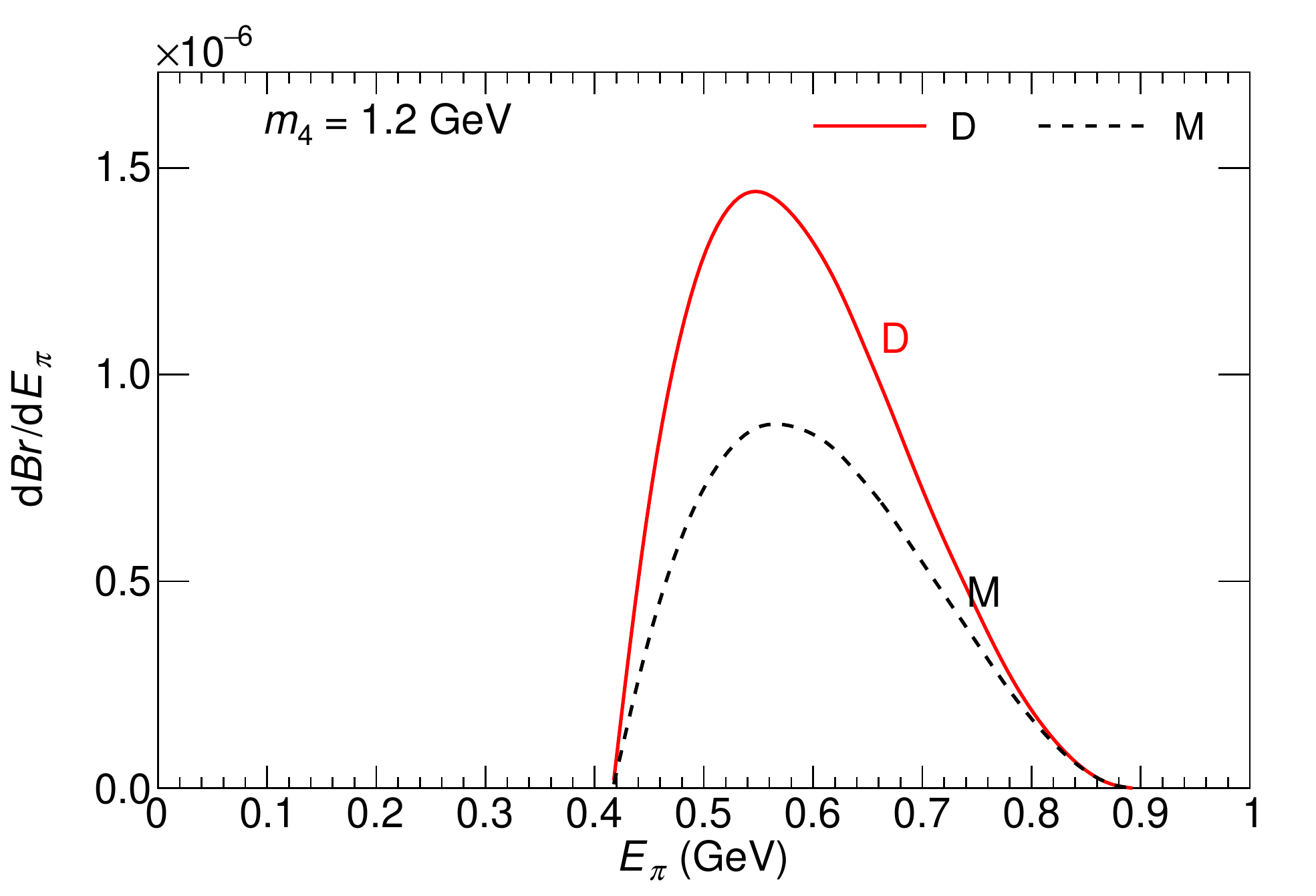}
   \label{d}
 }
   \subfigure[]{
   \includegraphics[scale=0.35] {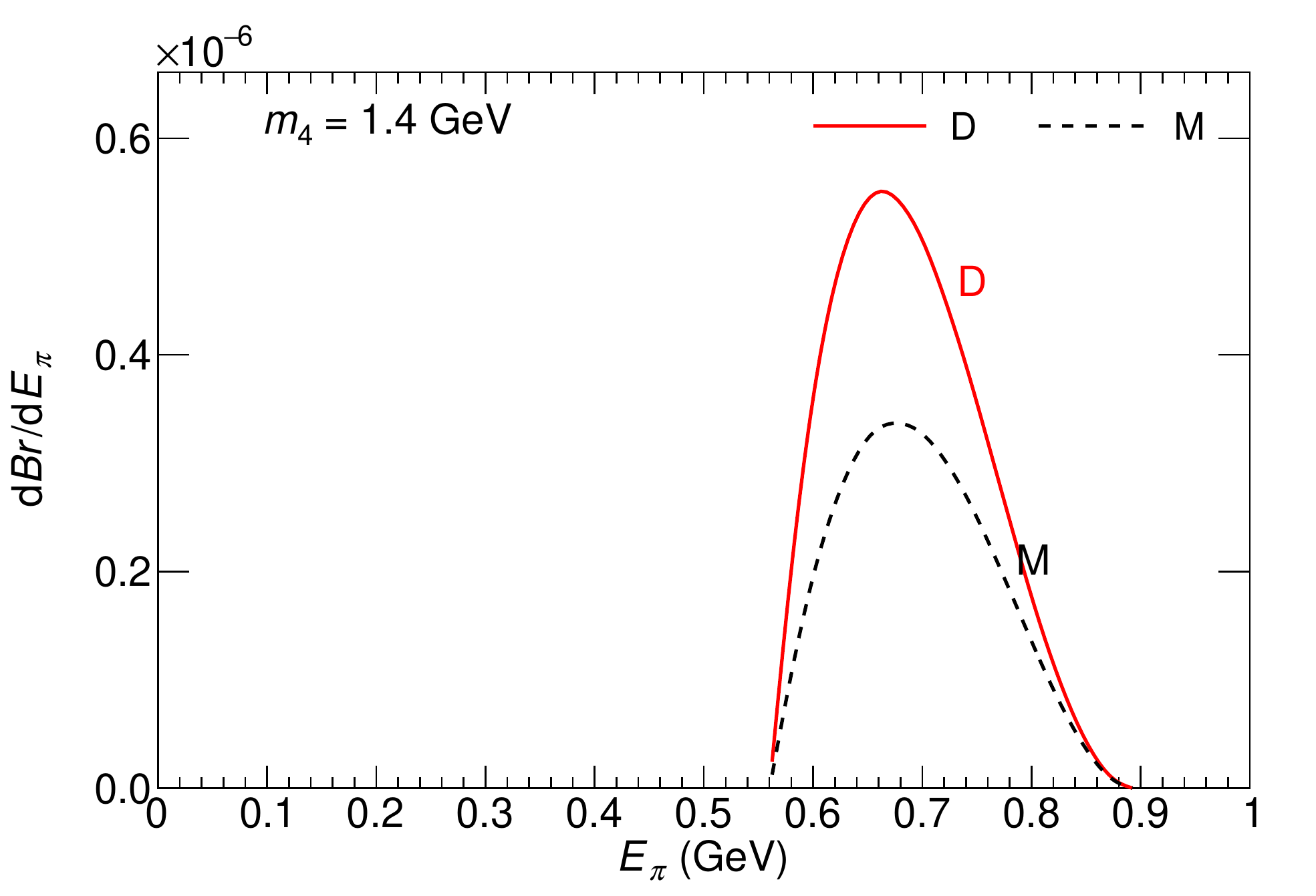}
   \label{e}
 }
    \subfigure[]{
   \includegraphics[scale=0.35] {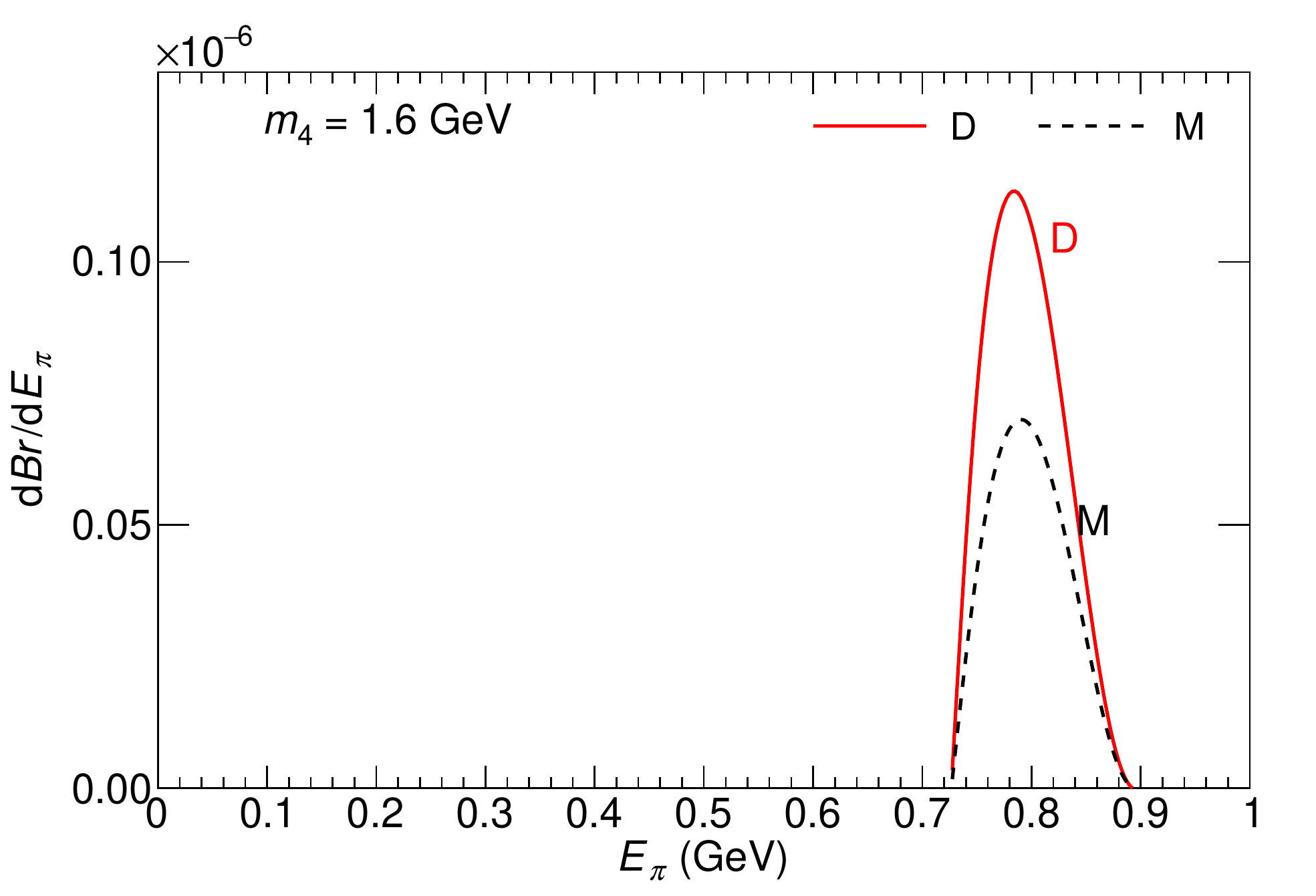}
   \label{f}
 }
\caption{The differential branching ratio $dBr/dE_\pi~(\mathrm{GeV}^{-1})$ as a function of $\pi$ energy in the $\tau$ rest frame with $|V_{eN}|^2=3.0\times10^{-3}$ and $|V_{\tau N}|^2=6.0\times10^{-3}$, for the decays $\tau^+ \to e^+ e^+ \pi^- \nu$, for different neutrino mass: \ref{a} $m_N =0.5~\mathrm{GeV}$; \ref{b} $m_N =0.8$ GeV; \ref{c} $m_N =1.0$ GeV; \ref{d} $m_N =1.2$ GeV; \ref{e} $m_N =1.4$ GeV; \ref{f} $m_N =1.6$ GeV. In each one there are two curves, corresponding to different type of neutrino. The red solid curve represents the case of Dirac neutrino and the black dash curve represents the Majorana neutrino.} \label{fig:result}
\end{figure}

\begin{figure}[ht] \label{resultmin}
\centering
\subfigure[]{
   \includegraphics[scale=0.35] {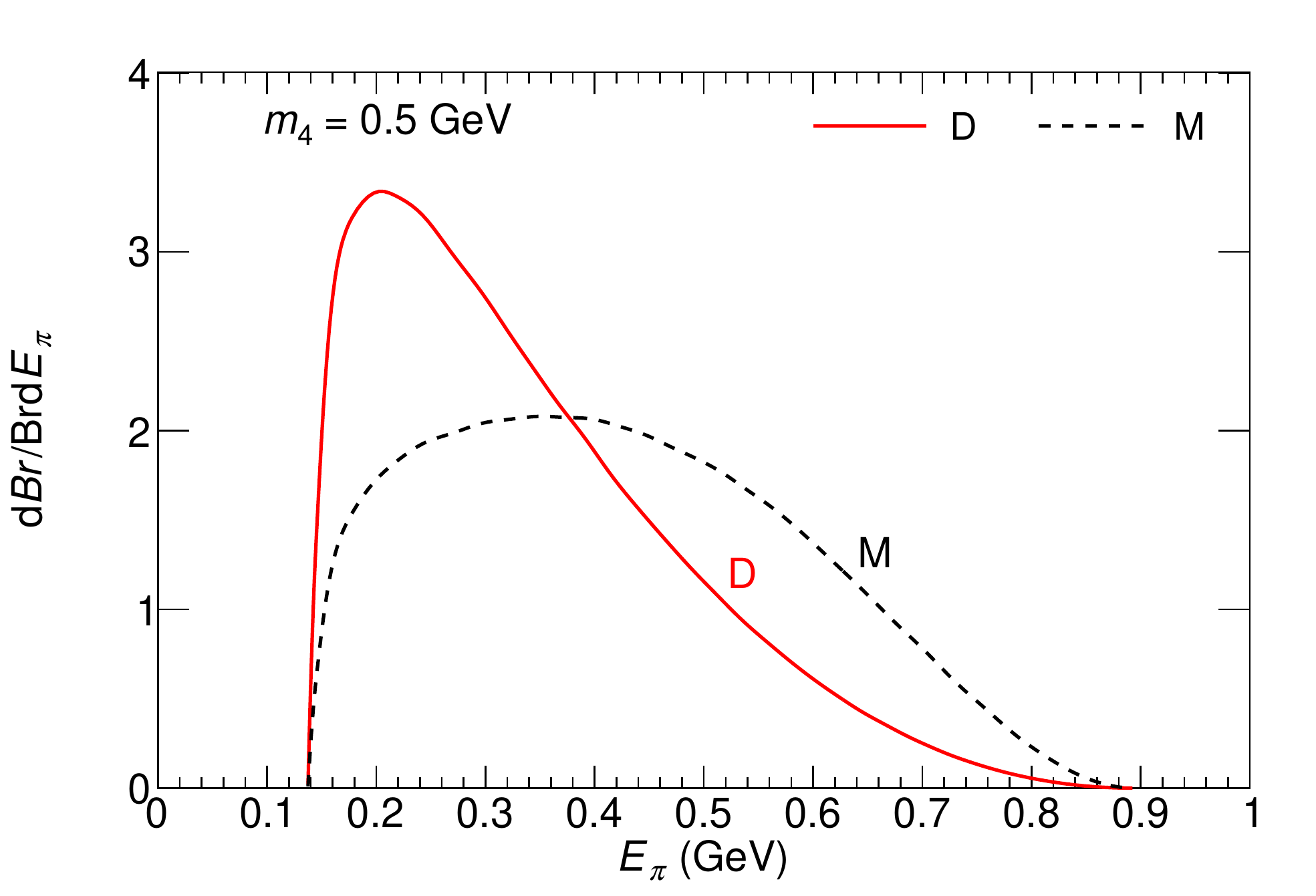}
   \label{amin}
 }
 \subfigure[]{
   \includegraphics[scale=0.35] {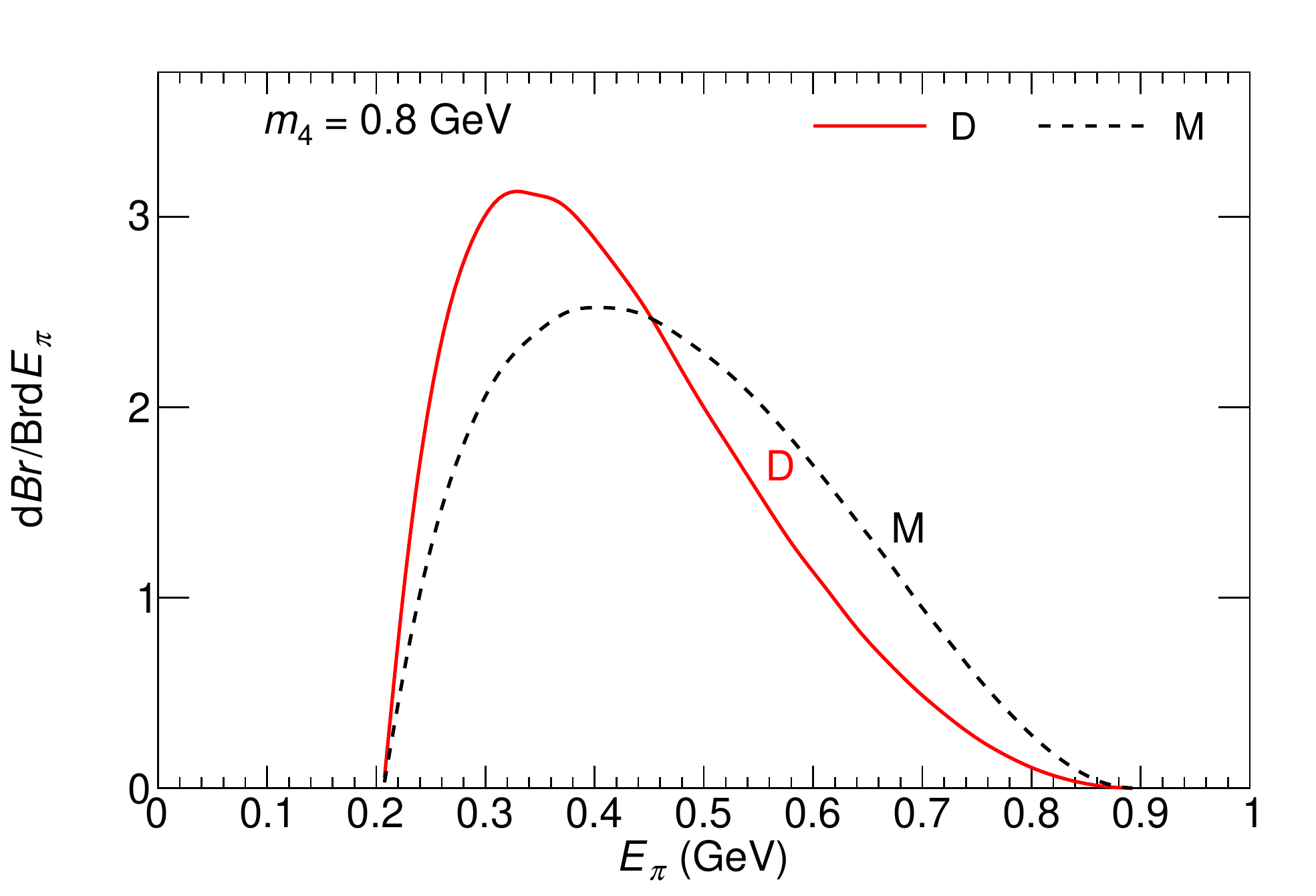}
   \label{bmin}
 }
 \subfigure[]{
   \includegraphics[scale=0.35] {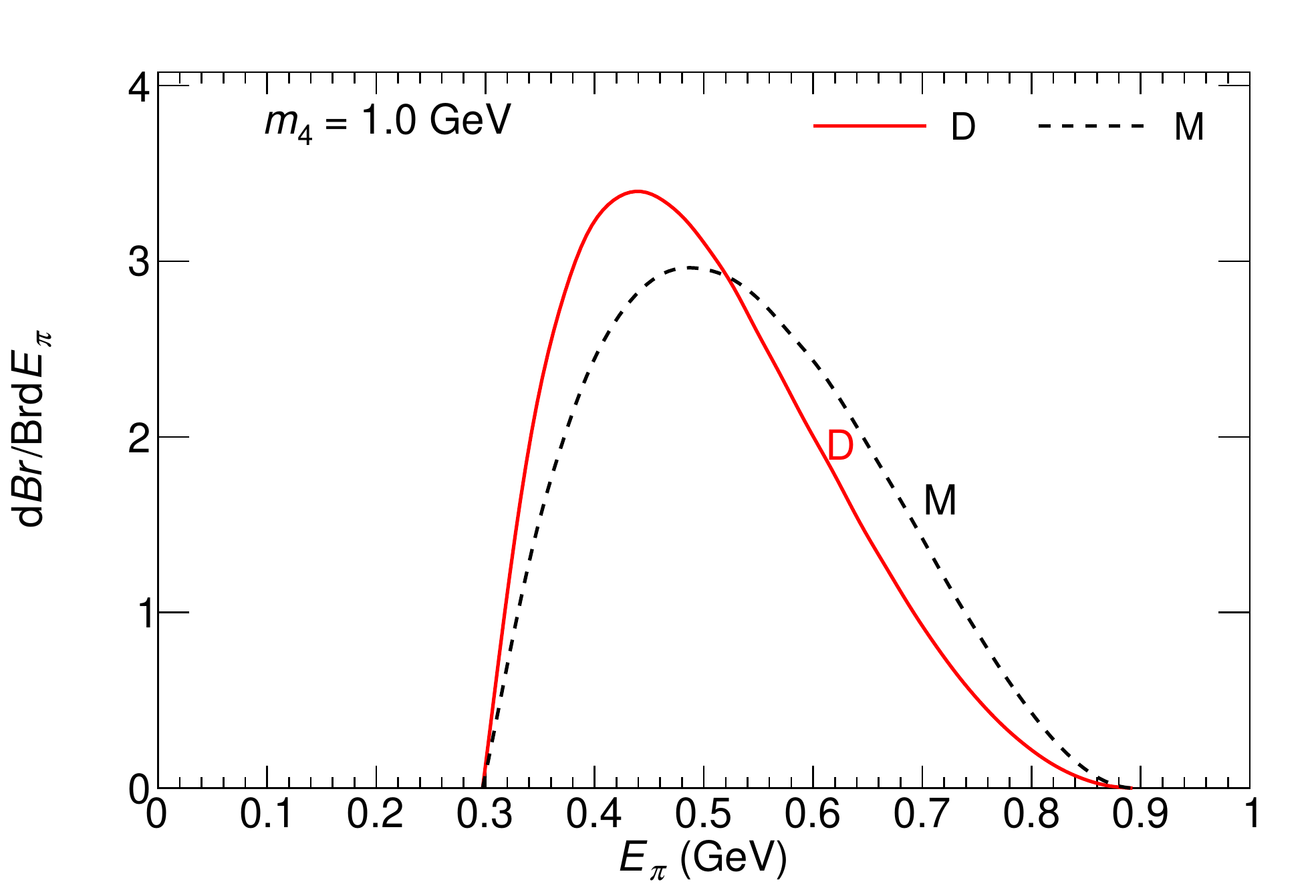}
   \label{cmin}
 }
  \subfigure[]{
   \includegraphics[scale=0.35] {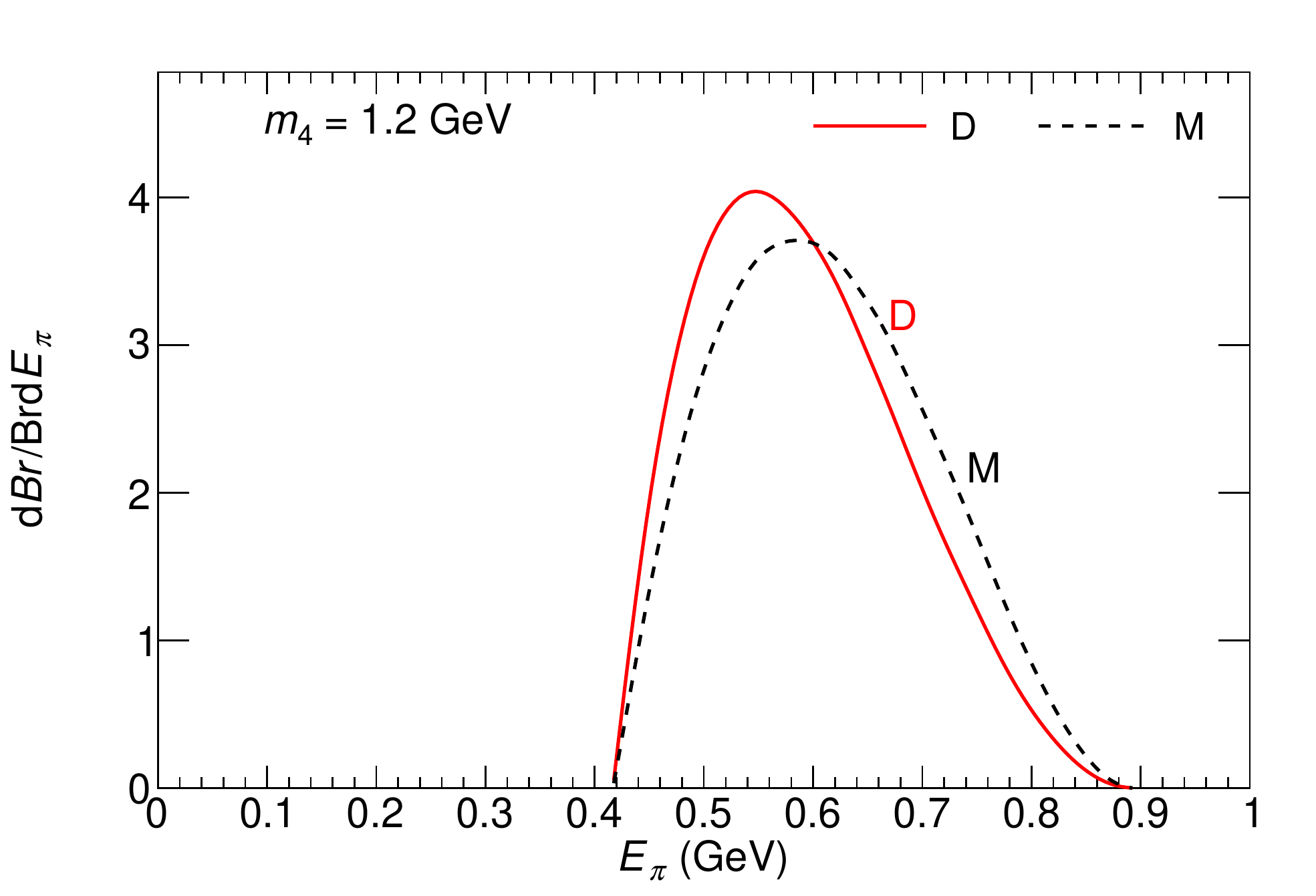}
   \label{dmin}
 }
   \subfigure[]{
   \includegraphics[scale=0.35] {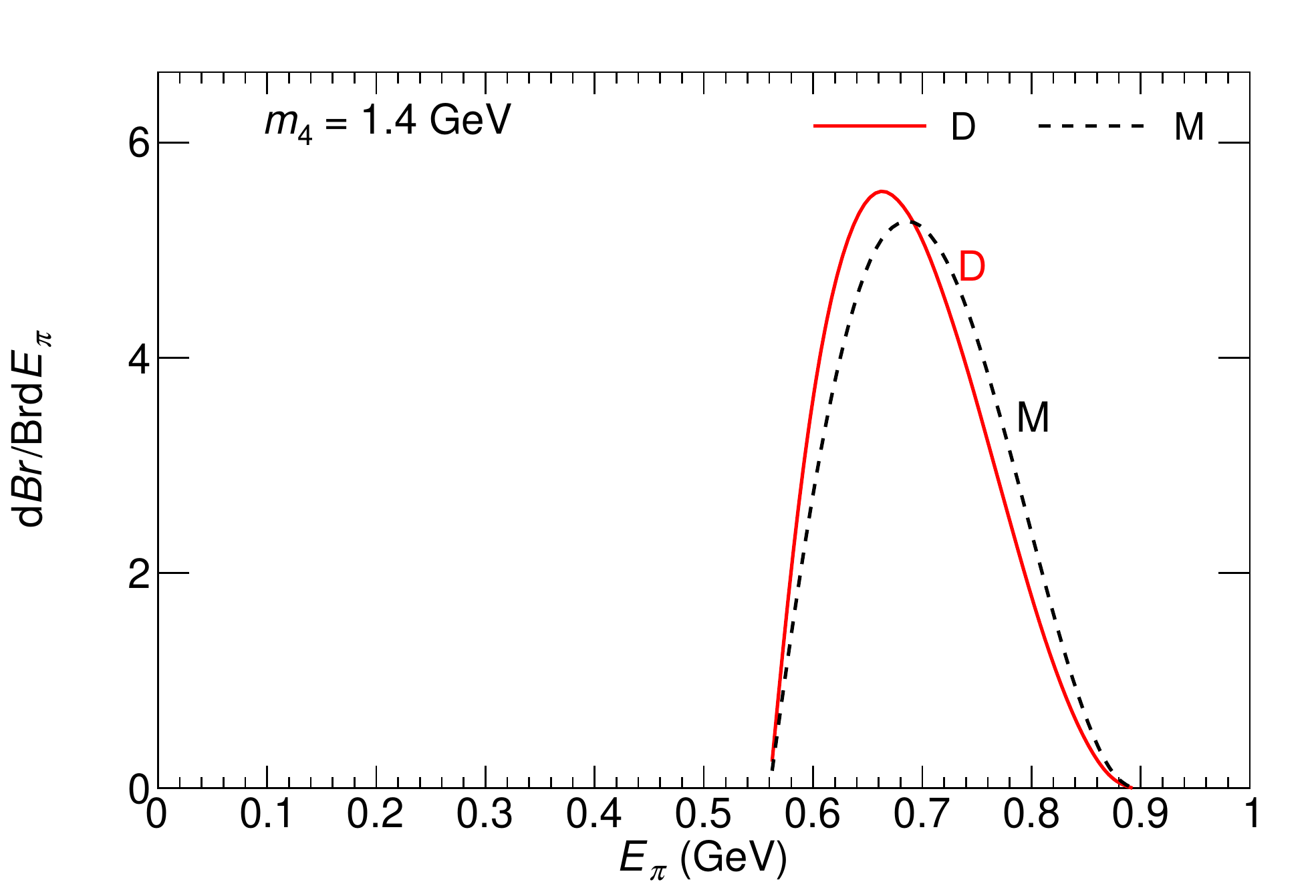}
   \label{emin}
 }
    \subfigure[]{
   \includegraphics[scale=0.35] {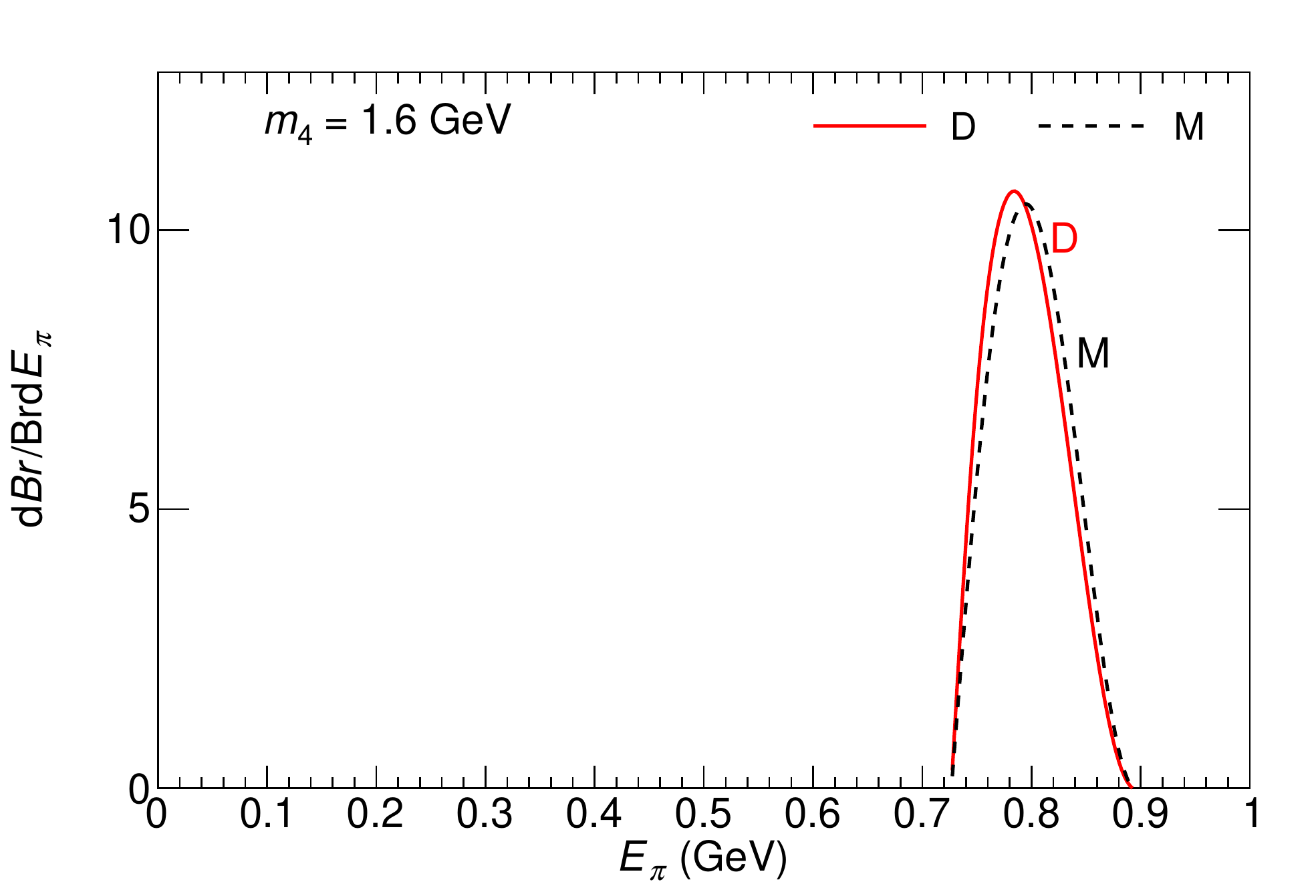}
   \label{fmin}
 }
\caption{As Figure 5, but for $|V_{eN}|^2=3.0\times10^{-3}$ and $|V_{\tau N}|^2=3.0\times10^{-3}$ and with normalized distributions.} \label{fig:resultmin}
\end{figure}

Physically, according to the energy and momentum conserving laws, and by the rebuilding of the vertexes, we can distinguish the two Feynman diagrams of Fig. 1 and Fig. 3. If there is only Fig. 3 exists, the exchanging heavy neutrino is Dirac neutrino; if both diagrams in Fig. 1 and Fig. 3 exist, the neutrino should be Majorana neutrino. This difference also has effects in the branching ratios and differential branching ratios. But since we do not know the exact mixing parameters for Majorana and Dirac neutrinos, the total branching ratio is not a good way to distinguish them. Nevertheless the differential branching ratios can be used to distinguish them. We calculate the differential branching ratio $dBr/dE_\pi$ of $\tau^+ \to e^+ e^+ \pi^- \bar{\nu}_\tau$ and $\tau^+ \to e^+ e^+ \pi^- \nu_e$ with $|V_{eN}|^2=3.0\times10^{-3}$ and $|V_{\tau N}|^2=6.0\times10^{-3}$, which are shown in Fig. \ref{fig:result}. And Fig. \ref{fig:resultmin} show the same value only with $|V_{eN}|^2=3.0\times10^{-3}$ and $|V_{\tau N}|^2=3.0\times10^{-3}$ and with normalized distributions.

In all Fig. \ref{fig:result} sub-figures, as the reason of mixing parameters, most Dirac cases curves are above Majorana cases. In these figures we are able to see the shape of differential branching ratios can distinguish the type of neutrino to a certain extent. Along with the increasing of neutrino mass the difference gets smaller and smaller. If the neutrino mass is 0.5 GeV, the disparity between these two curves is the largest. In Fig. \ref{a} with the growing of $E_\pi$, the trend of two curves are different. If the neutrino is Dirac neutrino, the differential branching ratio rises to maximum quickly around at $E_\pi=200~\mathrm{MeV}$ then it drops down. The extremum of the Majorana case appears also around at $E_\pi=200~\mathrm{MeV}$ then it decreases gently. From Fig. \ref{b} to Fig. \ref{f} the curve of Dirac neutrino almost cocoons the Majorana neutrino, which result in the difficulty to distinguish these two curves.

\begin{figure}[!h]
\centering
\includegraphics[scale=0.5]{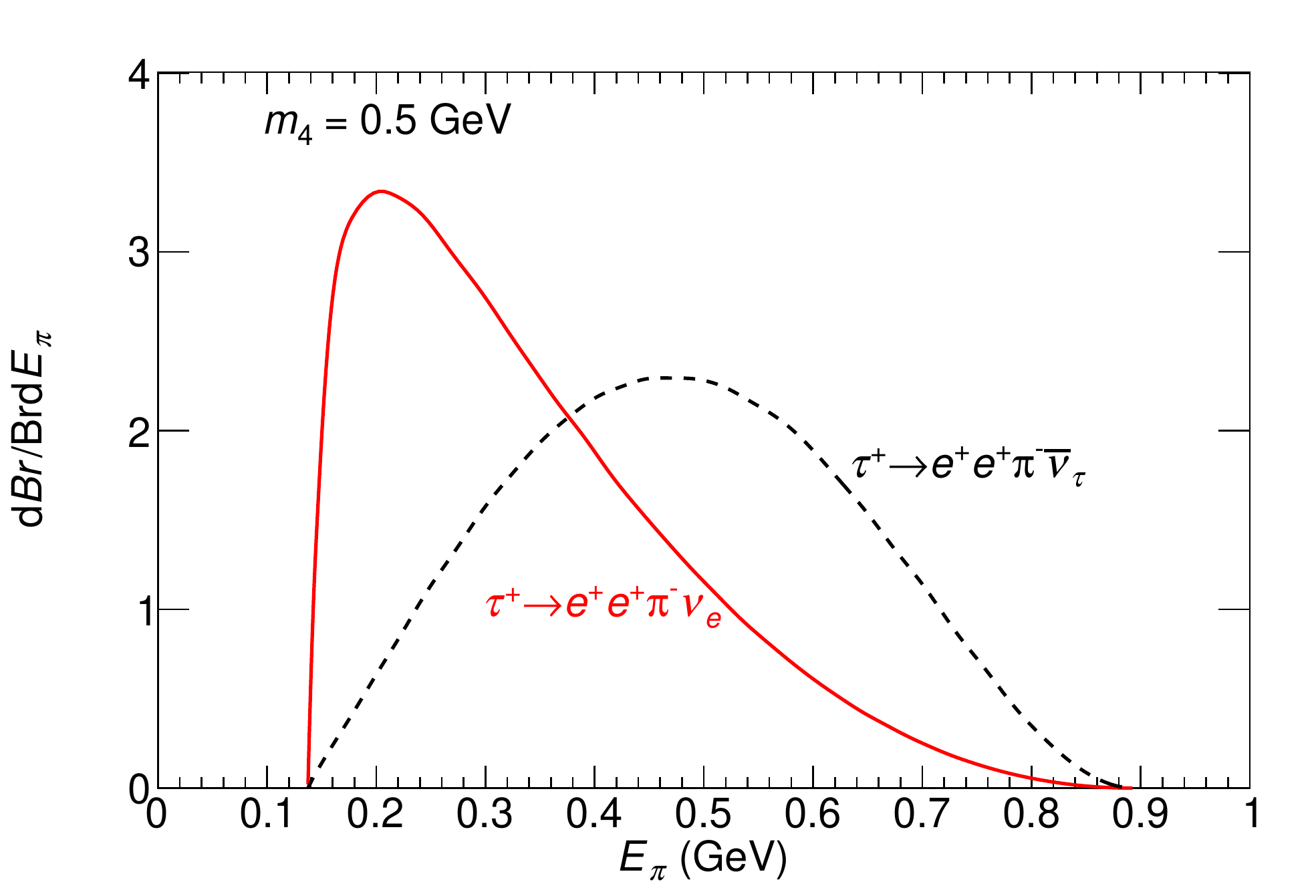}
\caption{The normalized differential branching ratio $dBr/BrdE_\pi(\mathrm{GeV}^{-1})$ as a function of $\pi$ energy in the $\tau$ rest frame with heavy Majorana neutrino. The red soild line represents $\tau^+\to e^+e^+\pi^-\nu_e$ and the black dash line is $\tau^+\to e^+e^+\pi^-\bar{\nu}_\tau$. \label{fig:dm5}}
\end{figure}

In Fig. \ref{fig:resultmin} the differential branching ratios are obtained of Majorana and Dirac cases with same mixing parameters $|V_{\ell4}|^2$. In sub-figure \ref{amin}, If the neutrino is Dirac neutrino, the differential branching ratio rises to maximum still around at $E_\pi=200~\mathrm{MeV}$ then it drops down quickly. But for Majorana case the maximum appears at $E_\pi=400~\mathrm{MeV}$, and the whole curve changes gently. In Fig. \ref{bmin} the difference between two curves grow smaller. Along with the increase of heavy neutrino mass, the distinction between two cases grows less. In Fig. \ref{emin} and \ref{fmin} the red solid line covers the black dash line. If $|V_{e4}|^2\gg|V_{\tau4}|^2$, the LNV process $\tau^+ \to e^+ e^+ \pi^- \bar{\nu}_\tau$ dominates, and spectrum will show more clearly its shape, representing the presence of a Majorana neutrino. On the other hand, if $|V_{e4}|^2\ll|V_{\tau4}|^2$, the LFV process $\tau^+ \to e^+ e^+ \pi^- \nu_e$ dominates, even if heavy neutrino is Majorana neutrino, the spectrum still show the same shape of a Dirac heavy neutrino. In $\tau$ rest frame with smaller heavy neutrino mass range, using $\pi$ spectrum to distinguish Majorana and Dirac Neutrino has good performance. While in larger heavy neutrino mass it does not work well. Since in Fig. \ref{amin} the difference between Majorana and Dirac cases is the largest. So we draw Fig. \ref{fig:dm5} to explore the deep reason.
Fig. \ref{fig:dm5} shows the normalized differential branching ratios $dBr/BrdE_\pi$ of $\tau^+\to e^+e^+\pi^-\nu_e$ and $\tau^+\to e^+e^+\pi^-\bar{\nu}_\tau$ with heavy Majorana neutrino mass $m_4=0.5~\mathrm{GeV}$. To obtain the normalized differential branching ratio of heavy Majorana neutrino, we need to add the red and black line together; as for Dirac case, it is two times of red line (since the decay width of Dirac neutrino is half of Majorana neutrino). In Fig. \ref{fig:dm5}, $\tau^+\to e^+e^+\pi^-\nu_e$ peaks more sharply (at a smaller energy), whereas $\tau^+\to e^+e^+\pi^-\bar{\nu}_\tau$ is flatter with a peak at a higher energy. So in Fig. \ref{amin}, the combination of $\tau^+\to e^+e^+\pi^-\nu_e$ and $\tau^+\to e^+e^+\pi^-\bar{\nu}_\tau$ (black dash line there) is gentler than red solid line there. It is the essential difference between $\tau^+\to e^+e^+\pi^-\nu_e$ and $\tau^+\to e^+e^+\pi^-\bar{\nu}_\tau$ that brings the distinction in Fig. \ref{fig:dm5}. And as mentioned before, different amplitudes of these two processes is definitely the most important one of all the possible reasons in our calculation.

Ref. \cite{Cvetic:2012hd,Cvetic:2015naa} use differential branching ratio $dBR/dE_{\mu}$ (muon energy distribution of the rare decay $\pi^+\to e^+e^+\mu^-\nu$) as a tool to distinguish between Dirac or Majorana neutrinos. In this paper we use differential branching ratio $dBR/dE_{\pi}$ to do the same thing. In the previous work the initial particle is $\pi$ meson, so the phase space of $\mu$ lepton is smaller than $\pi$. In our work, the initial particle is $\tau$, the final $\pi$ meson phase space gets larger than $\pi$ and smaller than $\tau$. So this work can be treated as a supplement to the previous work. Two final leptons in $\tau^+ \to e^+ e^+ \pi^- \bar{\nu}_\tau$ and $\tau^+ \to e^+ e^+ \pi^- \nu_e$ are the same. If we choose $dBR/dE_e$ to distinguish Majorana or Dirac Neutrino, it needs to be ensured that the $e$ leptons produced in the similar vertexes of these two processes. So it seems that $dBR/dE_\pi$ is a good choice. In $\tau^+ \to e^+ e^+ \pi^- \bar{\nu}_\tau$ and $\tau^+ \to e^+ e^+ \pi^- \nu_e$ processes, the final leptons can also be $\mu^+\mu^+$ or $e^+\mu^+$. Considering the $\mu^+$ and $e^+$ lepton share the same mixing parameters limits\cite{delAguila:2008pw} and $e^+$ lepton provides larger phase space for $\pi$ meson with the same heavy neutrino mass. So we choose $e^+e^+$ in the finial state as a representative instead of $\mu^+\mu^+$ and $e^+\mu^+$.

And we also need to consider the situation about experiment. The $\tau^+\tau^-$ cross section is $0.919~\mathrm{nb}$, giving $719~(430)$ million $\tau$ lepton pairs in the Belle (BaBar) data set. KEK and Belle-II upgrade program will ultimately yield a factor of 50 increase in integrated luminosity. The upgrade of the LHC accelerator and the LHCb detector will produce a data sample corresponding to an integrated luminosity of $50~\mathrm{fb}^{-1}$ \cite{Bediaga:2012py} at $\sqrt{s}$ of $13~\mathrm{TeV}$. Taking the ratio of $13~\mathrm{TeV}$ to $7~\mathrm{TeV}$ heavy-quark production cross section to be 1.8 \cite{Aaij:2011jh,Aaij:2015rla,Aaij:2013mga,Aaij:2015bpa}, the $\tau$ lepton yield will increase by approximately a factor of 30. The ATLAS expects $\tau$ lepton yields can be scaled to $3~\mathrm{ab}^{-1}$ with a factor of $1.6$ increase in cross section \cite{Chatrchyan:2014mua,Aad:2016naf}. Belle collaboration gives the $\tau$ lepton LNV processes as $Br\leq10^{-8}$ \cite{Miyazaki2013}. In theory, if we choose strict limits of mixing parameters $|V_{e4}|^2\sim10^{-8}$, which may lead to branching ratio $Br\leq10^{-8}$. Considering current experiment limits from Belle and BaBar, detecting these type LNV processes is still difficult. Future circular collider (FCC) \cite{Benedikt:2015kqj}, a proton-proton collider with $\sqrt{s}=100\mathrm{TeV}$ would have about seven times cross section for $W$ and $Z$ production than LHC. We may expect it can produce enough $\tau$ lepton events for searching $\tau$ LNV decays. Another challenging issue is the ununcertainty of $\pi$ meson. The determination of $\pi$ energy in the lab frame needs an uncertainty below $10$ MeV to achieve the requirement of discrimination. In ILC, whose $\delta E/E$ can reach $10^{-5}$\cite{Adolphsen2013} (which means that a $100~\mathrm{GeV}$ $\pi$ can be measured with a precision of a few times $10~\mathrm{MeV}$). If in future detector the $\pi$ meson energy satisfies this condition, the uncertainty is small enough for detecting.

\section{Summary and conclusions}
We choose $\tau$ lepton decays $\tau^+ \to e^+ e^+ \pi^- \bar{\nu}_\tau$ and $\tau^+ \to e^+ e^+ \pi^- \nu_e$ to determine the nature of neutrino. First, if either decay takes place, it means that there are heavy sterile neutrino exist. Second, basically we can distinguish these two decays by energy and momentum conserving laws. If both cases occur, the exchanging neutrino is Majorana neutrino; if only process $\tau^+ \to e^+ e^+ \pi^- \nu_e$ occur, the neutrino is Dirac neutrino. The nature of neutrino can also be determined by the differential branching ratio $dBr/dE_\pi$ in some extent. In our calculation, the internal exchanging heavy sterile neutrino is on mass-shell, which will enhance the decay rate by several orders and make the detection of these decays possible in current and near future experiment.

\section*{Acknowledgments}
We would like to thank Tao Han for his suggestions to carry out this research and providing the FORTRAN codes hanlib for the calculations. This work was supported in part by the National Natural Science Foundation of China (NSFC) under grant No.~11405037, 11575048 and 11505039.

\appendix
\renewcommand{\appendixname}{Appendix}
{\section{Calculation details of $\tau^+ \to e^+ e^+ \pi^- \bar{\nu}_\tau$ and $\tau^+ \to e^+ e^+ \pi^- \nu_e$}
In this appendix we present general formulas for thr LNV decay $\tau^+ \to e^+ e^+ \pi^- \bar{\nu}_\tau$ and LFV decay $\tau^+ \to e^+ e^+ \pi^- \nu_e$ in Fig. \ref{fig:feynM} and Fig. \ref{fig:feynMD}, respectively. Both decays are assumed to take place via the exchange of an on-shelll neutrino $N$. The transition amplitude of LNV process $\tau^+ \to e^+ e^+ \pi^- \bar{\nu}_\tau$ in Fig. \ref{fig:feynM} is in Eq. \eqref{eq:amplitudeM}. Since the process is dominated by on mass shell intermediate neutrino $N$, In the calculation of branching ratio it is reasonable to use narrow width approximation
\begin{equation}
\frac{1}{(q^2-m_4^2)^2+m_4^2\Gamma_{N_4}^2}\simeq\frac{\pi}{m_4\Gamma_{N_4}}\delta(q^2-m_4^2).
\end{equation}

For the calculation of decay width
\begin{equation}\label{eq:width}
\Gamma(\tau^+(P) \to e^+(P_2) e^+(P_3) \pi^-(P_4) \bar{\nu}_\tau(P_1))=\frac{1}{2m_\tau}\int d_{ps4}|\mathcal{M}|^2,
\end{equation}
where $d_{ps4}$ is the four-body phase spaces integration. The specific form is
\begin{equation}\label{eq:4body}
d_{ps4}(P\to P_1P_2P_3P_4)=\frac{d^3P_1}{(2\pi)^32E_1}\frac{d^3P_2}{(2\pi)^32E_2}\frac{d^3P_3}{(2\pi)^32E_3}\frac{d^3P_4}{(2\pi)^32E_4}(2\pi)^4\delta^4(P-P_1-P_2-P_3-P_4).
\end{equation}
The four-body phase spaces integral can be decomposed into three-body phase space integral $d_{ps3}(P\to P_1P_2q)$ and two-body phase spaces $d_{ps2}(q\to P_3P_4)$. Then the Eq. \eqref{eq:4body} can be written as
\begin{eqnarray}\label{eq:4to23}
d_{ps4}(P\to P_1P_2P_3P_4)&=&d_{ps3}(P\to P_1P_2q)\times \frac{dm_4^2}{2\pi} \times d_{ps2}(q\to P_3P_4)\nonumber\\
&=&\int \frac{dm_4^2}{2\pi}\int\frac{d^3P_1}{(2\pi)^32E_1}\frac{d^3P_2}{(2\pi)^32E_2}\frac{d^3q}{(2\pi)^32E_q}(2\pi)^4\delta^4(P-P_1-P_2-q)\nonumber\\
&&\times\int\frac{d^3P_3}{(2\pi)^32E_3}\frac{d^3P_4}{(2\pi)^32E_4}(2\pi)^4\delta^4(q-P_3-P_4),
\end{eqnarray}
where two-body phase $d_{ps2}$ is
\begin{equation}\label{eq:2body}
d_{ps2}(q\to P_3P_4)=\int\frac{1}{(2\pi)^6}\frac{\pi}{2m_4^2}\lambda^{\frac{1}{2}}(m_4^2,m_\ell^2,m_\pi^2)\frac{d\Omega}{4\pi},
\end{equation}
with $\lambda^{1/2}$ is the square root of the function
\begin{equation}
\lambda(x,y,z)\equiv x^2+t^2+z^2-2xy-2yz-2xz,
\end{equation}
and $d\Omega=d\cos\theta d\phi$. Since we use Monte Carlo method to get the integral value of decay width in this paper, so the $d\Omega$ can be rewritten like
\begin{equation}
\phi=2\pi x_1,\cos\theta=1-x_2,
\end{equation}
where $x_1,x_2$ is in range $(0\sim1)$. As for three-body phase spaces $d_{ps3}(P\to P_1P_2q)$ can be transformed as a chain of two-body phase spaces
\begin{equation}\label{eq:3body}
d_{ps3}(Y\to abc)=d_{ps2}(Y\to Xc)d X^2d_{ps2}(X\to ab).
\end{equation}
The chain is allowed for the following range of $X$, $a+b\leq X \leq Y-c$. With Eq. \eqref{eq:4body}, \eqref{eq:4to23}, \eqref{eq:2body} and \eqref{eq:3body}, formula \eqref{eq:width} full form is
\begin{eqnarray}\label{eq:lv2width}
\Gamma(\tau^+\to e^+e^+\pi^-\bar{\nu}_\tau)&=&\int\frac{G_F^4|V_{e4}V_{e4}|^2V_x^2f_x^2}{32m_{\tau}^3\pi^4M_{W}^2m_4\Gamma_{N4}}(2P\cdot P_4P_1\cdot P_2P_3\cdot P_4-m_x^2P\cdot P_3P_1\cdot P_2)\nonumber\\
&&\times\lambda^{\frac{1}{2}}(m_\tau,m_{\bar{\nu}_\tau}^2,m_W^2)\lambda^{\frac{1}{2}}(m_W,m_{e}^2,m_4^2)\lambda^{\frac{1}{2}}(m_4^2,m_{e}^2,m_\pi^2)\nonumber\\
&&\times dm_W^2dx_1dx_2dx_3dx_4dx_5dx_6.
\end{eqnarray}
Thus we can use VEGAS (Monte Carlo integral code) to calculate four-body phase spaces integral.
Then the branching ratio  $Br=\tau_\tau\Gamma(\tau^+\to e^+e^+\pi^-\bar{\nu}_\tau)$, where $\tau_\tau$ is the lifetime of $\tau$. $\Gamma(\tau^+\to e^+e^+\pi^-\nu_e)$ can be gotten in the same way.
\begin{eqnarray}\label{eq:lfwidth}
\Gamma(\tau^+\to e^+e^+\pi^-{\nu}_e)&=&\int\frac{G_F^4|V_{e4}V_{\tau4}|^2V_x^2f_x^2}{32m_{\tau}^3\pi^4M_{W}^2m_4^3\Gamma_{N4}}\left[4P\cdot P_1P_2\cdot P_3(P_3\cdot P_4)^2-m_e^2m_\pi^2P\cdot P_1P_2\cdot P_3\right.\nonumber\\
&&\left.-2m_e^2m_\pi^2P\cdot P_1P_2\cdot P_4-2m_e^2 P\cdot P_1P_2\cdot P_4P_3\cdot P_4\right.\nonumber\\
&&\left.+m_\pi^4 P\cdot P_1P_2\cdot P_3+4m_\pi^2 P\cdot P_1P_2\cdot P_3P_3\cdot P_4\right]\nonumber\\
&&\times\lambda^{\frac{1}{2}}(m_\tau,m_4^2,m_W^2)\lambda^{\frac{1}{2}}(m_W,m_{e}^2,m_{\nu_e}^2)\lambda^{\frac{1}{2}}(m_4^2,m_{e}^2,m_\pi^2)\nonumber\\
&&\times dm_W^2dx_1dx_2dx_3dx_4dx_5dx_6.
\end{eqnarray}

As the four-body phase spaces integral is complexity, we also use Monte Carlo method to get the differential branching ratio. In this work we aim to get differential branching ratio $dBr/dE_{\pi}$. We separate $E_\pi$ to several bins and record decay width value with $E_\pi$ in a specific bin. If the $E_\pi$ bin is narrow enough, the fraction of the decay width and size of $E_\pi$ bin can be treated as differential branching ratio $dBr/dE_{\pi}$. Fig. \ref{fig:result} and \ref{fig:resultmin} are both obtained in this way.
}

\bibliography{bibliography}
\end{document}